\documentclass[unsortedadress,preprint
 ,secnumarabic%
,amssymb, amsmath,nobibnotes,prd]{revtex4}
\begin{document}
\title{Noncommutative Topological Half-flat Gravity\footnote{This paper was prepared for a
special issue of {\it Gen. Rel. Grav.} in honor of Prof. Alberto Garc\'{\i}a on the occasion of
his 60th birthday.}}

\author{H. Garc\'{\i}a-Compe\'an}
\email{compean@fis.cinvestav.mx}
\affiliation{Departamento de F\'{\i}sica,
Centro de Investigaci\'on y de Estudios Avanzados del IPN\\
P.O. Box 14-740, 07000 M\'exico D.F., M\'exico}
\author{O. Obreg\'on}
\email{octavio@ifug3.ugto.mx}
\affiliation{Instituto de F\'{\i}sica de la Universidad de Guanajuato\\ P.O. Box E-143,
37150 Le\'on Gto., M\'exico}
\author{C. Ram\'{\i}rez}
\email{cramirez@fcfm.buap.mx}
\affiliation{Facultad de Ciencias F\'{\i}sico Matem\'aticas,\\
Universidad Aut\'onoma de Puebla, P.O. Box 1364, 72000 Puebla, M\'exico}

\date{\today}

\begin{abstract}

We formulate a noncommutative description of topological half-flat gravity in four
dimensions. BRST symmetry of this topological gravity is deformed through a 
twisting of the usual BRST quantization of noncommutative gauge theories. Finally it is
argued that resulting moduli space of instantons is characterized by the solutions 
of a noncommutative version of the Pleba\'nski's heavenly equation.

\end{abstract}
\vskip -1truecm
%\pacs{PACS numbers: }
\maketitle

\vskip -1.3truecm
\newpage

\setcounter{equation}{0}

%%%%%%%%%%%%%%%%%%%%%%%%%%%%%%%%%%%%%%%%%%%%%%%%%%%%%%%%%%%%%%%%%%%%%%%%%%%%%%%
%%%%%%%%%%%%%%%%%%%%%%%%%%%%%%%%%%%%%%%%%%%%%%%%%%%%%%%%%%%%%%%%%%%%%%%%%%%%%%%
\section{Introduction}

In recent years, renewed efforts have been performed in the formulation of a
noncommutative theory of gravitation, motivated in part by the understanding of the
short distance behavior of the
spacetime \cite{connesbook}. Some proposals based on
the recent developments are given in \cite{cham1,moffat,cham,cham2,zanon}. In
particular, in \cite{cham,cham2} a Seiberg-Witten map for the tetrad and the Lorentz
connection is given, where these fields were taken as components of a SO(4,1)
connection in the first work, and of a U(2,2) connection in the second one. In these
works a MacDowell-Mansouri (MM) type action was considered, invariant under the
subgroup U(1,1)$\times$U(1,1), and the excess of degrees of freedom, additional to
the ones of the commutative theory, is handled by means of constraints. In
\cite{zanon}, from the chosen constraints, a consistent noncommutative SO(3,1)
extension was proposed. In Ref. \cite{wess5} it has been shown that noncommutative
gauge theories, based on the Seiberg-Witten map, for any commutative theory invariant
under a gauge group $G$, can be constructed. The resulting noncommutative theory can
be regarded as an effective theory, invariant under the noncommutative enveloping
algebra transformations of $G$, which are induced by the commutative transformations of
$G$.

Following these developments, starting from a SL(2,{\bf C}) self-dual connection,
we have given a formulation for quadratic
noncommutative topological gravitation, which contains the SO(3,1) topological
invariants, namely the signature and Euler characteristic \cite{topologico}. In fact, the
noncommutative signature can be straightforwardly obtained, but the Euler invariant
cannot, as it involves the same difficulty as the MM action, which contains a
contraction with the Levi-Civita tensor, instead of the SO(3,1) trace. However, both
invariants can be combined into an expression given by the signature with a SO(3,1)
self-dual connection, which amounts to the SL(2,{\bf C}) signature.

Self-dual gravity (for a review, see \cite{reviewsdg}), from which the hamiltonian
Ashtekar's formulation \cite{ashtekar} can be obtained, has very useful properties
which have allowed the exploration of quantum gravity in the framework of loop
quantum gravity. In Ref. \cite{ncsdg} we have considered self-dual gravity in the
Pleba\'nski formulation \cite{pleban}, in order to make a proposal for a
noncommutative theory of gravity, which is fully invariant under the noncommutative
gauge transformations. Thus, Pleba\'nski formulation is written as a SL(2,{\bf C})
topological BF formulation (helicity formalism \cite{helicity}), given by the trace
of the two-form $B$, times the field strength \cite{gary}. The contact with Einstein
gravitation is done through the implementation of constraints on the $B$-field, which
are solved by the square of the tetrad one-form \cite{pleban}. This theory can be
restated in terms of self-dual SO(3,1) fields, the spin connection and the $B$-field.
After the identification of the $B$ two-form with the tetrad one-form squared, a
variation of this action with respect to the spin connection gives the vanishing of
the torsion. The resulting action contains Einstein gravitation plus an imaginary
term, which is identically zero due to the Bianchi identities. The noncommutative
version is obtained at the level of the SL(2,{\bf C}) theory, by the application of
the Moyal product.

On the other hand, topological gravity in four dimensions has been a very useful toy model, which
encodes some of the properties of quantum gravity. Among these theories, topological half-flat
gravity represents a cohomology theory of the moduli space ${\cal M}$ of the so called half-flat
metrics, which mean metrics with the self-dual part of Riemann tensor vanishing. In this theory,
the variables for the gravitational field consist of a self-dual two-form ${\cal S}$ and the
self-dual spin connection $\omega^+$. The solutions of the equations of motion (called in
\cite{pleban} as ${\cal S}_{\cal H}$-structure) for these variables determine the classical
moduli space of self-dual metrics ${\cal M}$. This structure was eventually used in \cite{kuni},
to propose a topological gravity theory describing the intersection theory on the moduli space
${\cal M}$. In the case of simply connected spacetimes, the self-dual spin connection can be
gauged away and fixed to zero. The remaining equations (${\cal S}'_{\cal H}$-structure) determine
then a moduli space ${\cal M}_{\cal H}$ generated by all solutions of the heavenly equation
\cite{jerzy}.

In the present paper we propose the noncommutative deformation of the topological half-flat
gravity proposed by Kunitomo in Ref. \cite{kuni}. In section 2 we give a brief overview of
topological gravity. Section 3 is devoted to review some aspects of Pleba\'nski action and ${\cal
S}_{\cal H}$-structures in self-dual gravity \cite{pleban,helicity}. In section 4 we formulate
the noncommutative deformation of the topological half-flat gravity. Finally, in section 5 we
give our final comments.

%%%%%%%%%%%%%%%%%%%%%%%%%%%%%%%%%%%%%%%%%%%%%%%%%%%%%%%%%%%%%%%%%%%%%%%%%%
%%%%%%%%%%%%%%%%%%%%%%%%%%%%%%%%%%%%%%%%%%%%%%%%%%%%%%%%%%%%%%%%%%%%%%%%%%
\section{Topological Gravity in Four Dimensions}

Originally the cohomological field theories were proposed as an interpretation of Donaldson
theory, describing the topology of four-dimensional manifolds, in terms of a suitable quantum
field theory \cite{topoym}. Topological gravity in four dimensions was a further proposal by
Witten in \cite{wtg} to construct a gravitational analog for Donaldson theory whose basic
variables
were the tetrad and the spin connection. Then some other extra fields were introduced in order to
have a theory with a fermionic BRST-like symmetry. The action is then written as a BRST
commutator and consequently it is a BRST invariant theory by construction.  In Refs.
\cite{brooks,labas}, Witten's Lagrangian for topological gravity was obtained from the suitable
chosen Lagrangian $\int_X d^4x \sqrt{g}\, C_{abcd} \widetilde{C}^{abcd}$, with $C_{abcd}$ the
Weyl tensor and $\widetilde{C}^{abcd}$ its Hodge dual, plus the gauge-fixing and ghost
Lagrangians. Witten's Lagrangian is then rederived through a genuine BRST procedure. The BRST
gauge-fixing procedure implies the introduction of new fields involving the gauge-fixing of
diffeomorphism, Weyl and Lorentz symmetries.

The moduli problem of topological gravity considered in \cite{wtg}, was the moduli space of
tetrads and spin connections satisfying the torsion free condition and the self-duality
condition of the Weyl tensor $\widetilde{C}^{abcd} = + {C}^{abcd}$. Further developments of
this proposal were considered in \cite{my,teo}. Of particular interest, among other moduli
problems, is the topological gravity theory based in the moduli of gravitational instantons
\cite{tgi}, where the moduli space is now based on the self-duality of the Riemann tensor
instead of the Weyl tensor. In these papers, the starting classical action which is BRST
gauge-fixed is a
linear combination of the Euler and the signature topological invariants of the
spacetime manifold $X$,
\begin{equation}
S= \int_X d^4x \sqrt{g} \bigg( A \varepsilon^{\mu \nu \rho \sigma}\varepsilon_{abcd} R^{ab}_{ \
\ \mu \nu} R^{cd}_{ \ \ \rho \sigma} + B \varepsilon^{\mu \nu \rho \sigma} R^{ab}_{ \ \ \mu \nu}
R_{ab\rho \sigma} \bigg),
\label{topoaction}
\end{equation}
where $A$ and $B$ are arbitrary constants and $R^{ab}_{ \ \ \mu \nu} = \partial_{\mu}
\omega^{ab}_{\nu} - \partial_{\nu} \omega^{ab}_{\mu} -i [\omega_{\mu},
\omega_{\nu}]^{ab}.$

In all these mentioned moduli problems involving topological gravity in four dimensions in the
Palatini formalism whose basic set of variables include the tetrad (or metric) as well as the
spin connection. As
mentioned in the introduction, it
is very difficult to construct a sensible noncommutative extension of gravity theories
involving these field variables, and which at the same time preserves all their symmetries.
Fortunately, there is a formulation of topological gravity theory based not on the tetrad and
in the spin connection, but in this case, the tetrad is changed by the self-dual two-form $B$
mentioned in the introduction. This
theory has been proposed in Ref. \cite{kuni}, where it was called topological half-flat
gravity. In the present paper we will construct the noncommutative extension of the BRST
symmetry of this formulation, following the procedure of Ref. \cite{ncbrst}.

%%%%%%%%%%%%%%%%%%%%%%%%%%%%%%%%%%%%%%%%%%%%%%
%%%%%%%%%%%%%%%%%%%%%%%%%%%%%%%%%%%%%%%%%%%%%%
\section{Self-dual Variables}

One of the main features of the tetrad formalism of the theory of gravitation, is that it
introduces local Lorentz SO(3,1) transformations. In this case, the generalized Hilbert-
Palatini formulation is written as $\int e_a^{\ \mu}e_b^{\ \nu}R_{\mu\nu}^{\ \
ab}(\omega)d^4x$, where $e_a^{\ \mu}$ is the inverse tetrad, and $R_{\mu\nu}^{\ \ ab}(\omega)$
is the so(3,1) valued field strength. The decomposition of the Lorentz group as
SO(3,1)=SL(2,{\bf C})$\otimes$SL(2,{\bf C}), and the geometrical structure of four dimensional
space-time, makes it possible to formulate gravitation as a complex theory, as in
\cite{pleban,helicity}. These formulations take advantage of the properties of the fundamental
or spinorial representation of SL(2,{\bf C}), which allows a simple separation of the action
on the fields of both factors of SO(3,1), as shown in great detail in \cite{pleban,helicity}.
All the Lorentz Lie algebra valued quantities, in particular the connection and the field
strength, decompose into the self-dual and anti-self-dual parts, in the same way as the Lie
algebra so(3,1)=s$\ell$(2,{\bf C})$\oplus$s$\ell$(2,{\bf C}). However, Lorentz vectors, like
the tetrad, transform under mixed transformations of both factors and so this formulation
cannot be written as a chiral SL(2,{\bf C}) theory. Various proposals in this direction have
been made (for a review, see \cite{reviewsdg}).  In an early formulation, this problem has
been solved by Pleba\'nski \cite{pleban}, where by means of a constrained Lie algebra valued
two-form ${\cal S}$, the theory can be formulated as a chiral SL(2,{\bf C}) invariant
BF-theory, Tr$\int {\cal S}\wedge R(\omega)$. In this formulation ${\cal S}$ has two SL(2,{\bf
C}) spinorial indices, and it is symmetric on them ${\cal S}^{AB}={\cal S}^{BA}$, as any such
s$\ell$(2,{\bf C}) valued quantity. The constraints are given by ${\cal S}^{AB}\wedge {\cal
S}^{CD}=\frac{1}{3}\delta_{(A}^C\delta_{B)}^D{\cal S}^{EF}\wedge {\cal S}_{EF}$ and, as shown
in \cite{pleban}, their solution implies the existence of a tetrad one-form, which squared
gives the two-form ${\cal S}$. In the language of SO(3,1), this two-form is a second rank
antisymmetric self-dual two-form, ${\cal S}^{+ab}=\Pi^{+ab}_{\ \ \ cd}{\cal S}^{cd}$, where
$\Pi^{+ab}_{\ \ \ cd}=\frac{1}{4}\left(\delta_{cd}^{ab}-i\varepsilon^{ab}_{\ \ cd}\right)$. In
this case, the constraints can be recast into the equivalent form ${\cal S}^{+ab}\wedge {\cal
S}^{+cd}=- {1/3}\Pi^{+abcd}{\cal S}^{+ef}\wedge {\cal S}^{+}_{\ ef}$, with solution
${\cal S}^{ab}=2e^{a}\wedge e^{b}$.

For the purpose of the noncommutative formulation, we will consider self-dual gravity in a
somewhat different way as in the papers \cite{pleban,helicity}. In this section we will fix
our notations and conventions.

Let us take the self-dual SO(3,1) BF action, defined on a $(3+1)$-dimensional pseudo-riemannian manifold
$(X, g_{\mu \nu})$,
\begin{equation}
I=i{\rm Tr}\int_X {\cal S}^+\wedge R^+=i \int_X \varepsilon ^{\mu \nu \rho \sigma }
{\cal S}_{\ \mu \nu }^{+ \ ab}R_{\rho \sigma ab}^{+}(\omega )d^4x,\label{accion}
\end{equation}
where
$R_{\rho \sigma ab}^+=\Pi _{ab}^{+cd}R_{\rho \sigma cd}$,
is the self-dual SO(3,1) field strength tensor.
This action can be rewritten as
\begin{equation}
I=\frac{1}{2}\int_X \varepsilon^{\mu\nu\rho\sigma}\left(i{\cal S}_{\mu\nu}^{\ \ ab}R_{\rho
\sigma ab}+
\frac{1}{2}\varepsilon_{abcd}{\cal S}_{\mu\nu}^{\ \ ab}R_{\rho \sigma}^{\ \ cd}\right)d^4x.
\end{equation}
If now we take the solution of the constraints on ${\cal S}$, which we now write as
\begin{equation}
{\cal S}_{\mu\nu}^{\ \ ab}=e_\mu^{\ a}e_\nu^{\ b}-e_\mu^{\ b}e_\nu^{\ a},\label{sigma}
\end{equation}
then
\begin{equation}
I=\int_X (\det e R+i\varepsilon^{\mu\nu\rho\sigma}R_{\mu\nu\rho\sigma})d^4x.\label{accion1}
\end{equation}
The real and imaginary parts of this action must be variated independently
because the fields are real.
The first part represents Einstein action in the
Palatini formalism, from which, after variation of the Lorentz connection, a vanishing torsion
$T_{\mu\nu}^{\ \ a}=0$ turns out.
As a consequence, the second term vanishes due to Bianchi identities.

The action (\ref{accion}) can be written as
\begin{equation}
I=i\int_X \varepsilon ^{\mu \nu \rho \sigma }
{\cal S}_{\mu \nu }^{+\ ab}R_{\rho \sigma ab}(\omega^{+})d^4x,\label{accion3}
\end{equation}
where
$R_{\mu \nu }^{\ \ ab}(\omega^+)=
\partial _{\mu }\omega _{\nu }^{+ab}-\partial _{\nu}\omega _{\mu }^{+ab}+
\omega _{\mu }^{+ac}\omega _{\nu c}^{+b}-\omega _{\nu}^{+ac}\omega _{\mu c}^{+b}$.
From the decomposition SO(3,1)=SL(2,{\bf C})$\times$SL(2,{\bf C}), it turns out that
$\omega _{\mu }^{\ i}=\omega _{\mu}^{+0i}$ is a SL(2,{\bf C}) connection. Further, if we take
into account self-duality, $\varepsilon _{cd}^{\ \ ab}\omega _{\mu }^{+cd}=2i\omega _{\mu }^{+ab}$,
we get $\omega _{\mu }^{+ij}=-i\varepsilon _{\ \,k}^{ij}\omega _{\mu }^{\ k}$.
Therefore,
\begin{eqnarray}
R_{\mu \nu }^{\ \ 0i}(\omega ^{+}) &=&\partial _{\mu }\omega _{\nu
}^{i}-\partial _{\nu }\omega _{\mu }^{i}+2i\varepsilon _{jk}^{i}\omega _{\mu
}^{j}\omega _{\nu}^{k}=R_{\mu \nu }^{\ \ i}(\omega ), \\
R_{\mu \nu }^{\ \ ij}(\omega ^{+}) &=&\partial_\mu\omega_\nu^{+ij}-\partial_\nu\omega_\mu^{+ij}+
2(\omega_\mu^{\ i}\omega_\nu^{\ j}-\omega_\nu^{\ i}\omega_\mu^{\ j})=
-i{\varepsilon ^{ij}}_kR_{\mu \nu}^{\ \ k}(\omega ),
\end{eqnarray}
where ${R_{\mu\nu}}^i$ is the SL(2,{\bf C}) field strength.

Similarly, we define ${\cal S}_{\mu \nu }^{\ \ i}= {\cal S}_{\mu \nu }^{+\,0i}$, which
transforms
in
the SL(2,{\bf C}) adjoint representation. From it we get,
${\cal S}_{\mu \nu }^{+\,ij}=-i\varepsilon_{\ \,k}^{ij}{\cal S}_{\mu \nu }^{\ \ k}$. Thus,
the action (\ref{accion3}) can be written as a SL(2,{\bf C}) BF-action
$$
I=i\int_X \varepsilon ^{\mu \nu \rho \sigma }\left[ {\cal S}_{\mu \nu}^{+\,0i}
R_{\rho \sigma 0i}(\omega ^{+})+
{\cal S}_{\mu \nu }^{+\,ij}R_{\rho\sigma ij}(\omega ^{+})\right]d^4x
$$
\begin{equation}
=-4i\int_X \varepsilon ^{\mu \nu\rho \sigma }{\cal S}_{\mu \nu }^{\ \ i}
R_{\rho \sigma i}(\omega )d^4x.
\end{equation}
Therefore, if we choose the algebra $s\ell(2,{\bf C})$ to satisfy $[T_i,T_j]=-2\varepsilon_{ij}^{\ \,k}T_k$
and $Tr(T_iT_j)=-2\delta_{ij}$, we have that (\ref{accion}) can be rewritten as the self-dual action
\cite{pleban},
\begin{equation}
I=2i{\rm Tr}\int_X {\cal S} \wedge R = 2i \int_X {\cal S}_i \wedge R^i,
\label{accion4}
\end{equation}
which is invariant under the SL(2,{\bf C}) transformations
$\delta_\lambda\omega_\mu=\partial_\mu\lambda+i[\lambda,\omega_\mu]$ and $\delta_\lambda
{\cal S}_{\mu\nu}=i[\lambda,{\cal S}_{\mu\nu}]$.

If the variation of this action with respect to the SL(2,{\bf C}) connection $\omega$
is set to zero, we get the equations
\begin{equation}
\varepsilon ^{\mu \nu \rho \sigma }D_{\nu }{\cal S}_{\rho \sigma }^{\ \ i}=
\varepsilon ^{\mu \nu \rho \sigma }\left( \partial
_{\nu }{\cal S}_{\rho \sigma }^{\ \ i}+2i\varepsilon _{\ jk}^{i}\omega _{\nu }^{\ j}{\cal S}
_{\rho \sigma }^{\ \ k}\right) =0.
\end{equation}
Taking into account separately both real and imaginary parts, we get,
in terms of the SO(3,1) connection,
\begin{equation}
\varepsilon^{\mu\nu\rho\sigma}D_\nu {\cal S}_{\rho\sigma}^{\ \
ab}=\varepsilon^{\mu\nu\rho\sigma}\left(\partial_\nu {\cal S}_{\rho\sigma}^{\ \ ab}+
\omega_\nu^{\ ac}{\cal S}_{\rho\sigma c}^{\ \ \ b}-\omega_\nu^{\ bc} {\cal S}_{\rho\sigma c}^{\
\
\ a}
\right)=0,\label{constriccion}
\end{equation}
which after the identification (\ref{sigma}), can be written as
\begin{equation}
\varepsilon^{\mu\nu\rho\sigma}(\partial_\nu e_\rho^{\ a}e_\sigma^{\ b}-
\partial_\nu e_\rho^{\ b}e_\sigma^{\ a}
+\omega_\nu^{\ ac}e_{\rho c}e_\sigma^{\ b}-\omega_\nu^{\ bc}e_{\rho c}e_\sigma^{\ a})=
\varepsilon^{\mu\nu\rho\sigma}(T_{\nu\rho}^{\ \ a}e_\sigma^{\ b}-T_{\nu\rho}^{\ \ b}e_\sigma^{\ a})=0.
\end{equation}
From which the vanishing torsion condition once more turns out.

The underlying self-dual structure encoded in (\ref{accion4}), including constraints, can be
written in terms of the Lagrangian $L({\cal S}_{\cal H})$ given by,
\begin{equation}
L({\cal S}_{\cal H}):= \int_X {\cal S}_i \wedge R^i + \frac{1}{2} C_{ij} \big(\frac{1}{2} {\cal
S}^i \wedge {\cal S}^j - v \delta^{ij}\big).
\end{equation}
Similarly for the anti-self-dual structure, a Lagrangian $L({\cal S}_{{\cal H}'})$ can be defined
\cite{pleban,helicity}. Here $C_{ij}$ is the self-dual Weyl tensor, which is symmetric and
traceless. Further,  $v$ is the volume form given by $v=e^0 \wedge e^1 \wedge e^2 \wedge
e^3$ and $\delta_{ij} = {\rm diag(1,1,1)}$. 

Equations of motion and constraints can be written in the following way:
\begin{equation}
\delta {\omega}^i: \ \ \ \ \ \ \  D{\cal S}^i := d{\cal S}^i +  \varepsilon^{ijk}
\omega^j  \wedge {\cal S}^k = 0,\label{estruno}
\end{equation}
\begin{equation}
\delta {\cal S}^i: \ \ \ \ \ \ \ {\cal R}^i:= d\omega^i + \frac{1}{2}  \varepsilon^{ijk}
\omega^j  \wedge \omega^k=0,
\label{estrdos}
\end{equation}
\begin{equation}
\delta C_{ij}: \ \ \ \ \ \ \ \frac{1}{2} {\cal S}^i \wedge {\cal S}^j = \delta^{ij} \cdot v.
\label{estrtres}
\end{equation}
The Eqs. (\ref{estruno}), (\ref{estrdos}) and (\ref{estrtres})
define a self-dual Einstenian substructure ${\cal S}_{\cal H}$ of the complete structure
${\cal V}$ composed by the combinations of self-dual and anti-self-dual substructures, {\it i.e.}
${\cal
V}:= {\cal S}_{\cal H} \cup {\cal S}_{{\cal H}'}$, which are discussed in detail in
Refs. \cite{pleban,helicity}.

%%%%%%%%%%%%%%%%%%%%%%%%%%%%%%%%%%%%%%%%%%%%%%%%%%%%%%%%%%%%%%%%%%%%%%%%%%
%%%%%%%%%%%%%%%%%%%%%%%%%%%%%%%%%%%%%%%%%%%%%%%%%%%%%%%%%%%%%%%%%%%%%%%%%%
\section{Noncommutative Topological Half-flat Gravity}

Topological half-flat gravity is based on the moduli space ${\cal M}$ given by the solutions
to the ${\cal S}_{\cal H}$ structure. In this section we will consider a noncommutative
deformation of it. Thus the basic field variables are the noncommutative self-dual spin
connection ${\omega}^i$ and the noncommutative self-dual two-form ${\cal S}^i$ \cite{kuni},
defined on the noncommutative spacetime $\widehat{\cal A}_*(X) \equiv X_{\theta}$. 

The Lagrangian must be invariant under a noncommutative BRST-like transformation, $\delta_{Q}
\widehat{\cal O} = -i\{Q\stackrel{\ast}{,} \widehat{\cal O} \}$, where $\widehat{\cal O}$ is a
functional of the noncommutative fields, $Q$ is the BRST charge, and $\{\cdot \stackrel{\ast}{,}
\cdot \}$ is the BRST noncommutative commutator \cite{topoym,wtg}.

We will consider the fundamental BRST multiplet
$(\widehat{\omega}^I,\widehat{\cal S}^I,
\widehat{\psi}^I,
\widehat{\Psi}^I; \widehat{\gamma}^I,
\widehat{\gamma}^{\mu})$, which consists of the bosonic and ghost fields. As usual in
noncommutative theories, all fields are promoted to be valued on the enveloping algebra of
the adjoint representation ${\bf ad}$ of s${\ell}$(2,{\bf C}), that is on ${\cal
U}(s{\ell}(2,{\bf C}),{\bf ad})$. Indices $I,J,$ etc. run over all possible generators of
${\cal U}(s{\ell}(2,{\bf C}),{\bf ad})$. Here  $\widehat{\omega}^I$ is the spin connection
one-form, $\widehat{\psi}^I$ is a ghost one-form field associated
by the BRST symmetry  to $\widehat{\omega}^I$, $\widehat{\Psi}^I$ is a ghost
two-form field,  and $\widehat{\gamma}^I$ and $\widehat{\gamma}^{\mu}$ are the ghost fields
useful to fix the Lorentz and diffeomorphism symmetries respectively. 

The equations of the noncommutative self-dual $\widehat{\cal S}_{\cal H}$-structure are
given by:
\begin{equation}
\delta \widehat{\omega}^I: \ \ \ \ \ \ \  D\widehat{\cal S}^I := d\widehat{\cal S}^I + \frac{1}{2}
h^{IJK} \widehat{\omega}^J
\buildrel{*}\over{\wedge} \widehat{\cal S}^K = 0,
\label{ncestruno}
\end{equation}
\begin{equation}
\delta \widehat{\cal S}^I: \ \ \ \ \ \ \  \widehat{\cal R}^I:= d\widehat{\omega}^I + h^{IJK}
\widehat{\omega}^J
\buildrel{*}\over{\wedge} \widehat{\omega}^K=0.
\label{ncestrdos}
\end{equation}
Solutions of these equations
constitute the moduli space $\widehat{\cal M}$ of noncommutative `gravitational instantons'.

In local coordinates of $X_{\theta}$, under the assumption $h^{IJK} t^I = t^J \cdot t^K$, the
curvature two form is given by
$$
\widehat{\cal R}^I_{\mu \nu} = \partial_{\mu}\widehat{\omega}^I_{\nu}-
\partial_{\nu} \widehat{\omega}^I_{\mu} - i
[\widehat{\omega}_{\mu}\stackrel{\ast}{,}\widehat{\omega}_{\nu}]^I  =
\partial_{\mu}\widehat{\omega}^I_{\nu}-
\partial_{\nu} \widehat{\omega}^I_{\mu} - i
h^{IJK}\widehat{\omega}_{\mu}^J * \widehat{\omega}_{\nu}^K + i h^{IKJ}\widehat{\omega}_{\nu}^K *
\widehat{\omega}_{\mu}^J
$$
\begin{equation}
= \partial _{\mu}\widehat{\omega}^I_{\nu}- \partial_{\nu}\widehat{\omega}^I_{\mu} + \frac{1}{2}
f^{IJK} \left\{\widehat{\omega}^J_{\mu}\stackrel{\ast}{,}\widehat{\omega}^K_{\nu}\right\} -
\frac{i}{2} d^{IJK}
\left[
\widehat{\omega}^J_{\mu}\stackrel{\ast}{,}\widehat{\omega}_{\nu}^{K}\right],
\end{equation}
where $h^{IJK} = \frac{i}{2} f^{IJK} + \frac{1}{2} d^{IJK}$, with $d^{IJK}= d^{(IJK)}$. Here
$\left[A\stackrel{\ast}{,}B\right]\equiv
A\ast B-B\ast A$, $\left\{A\stackrel{\ast}{,}B\right\}\equiv
A\ast B+B\ast A$, and $*$ is given by the external
product of the Moyal product and the matrix
multiplication. Also $\left[ t^{I},t^{J}\right]=if^{IJK}t^{K}$, $\left\{
t^{I},t^{J}\right\}
= d^{IJK}t^{K}$ and the covariant derivative is given by $D_{\mu} \Theta(x) = \partial_{\mu} \Theta
+ i \left[\Theta \stackrel{\ast}{,}\omega_{\mu}\right]$, for any $\Theta$.

Before the full constraints are implemented, actually we do not have still the full BRST
symmetry.
Thus this preliminary BRST symmetry is simply the $S$-symmetry from Ref. \cite{kuni}. Thus the 
infinitesimal transformations for bosonic and ghost fields, now denoted by $\delta_S
\widehat{\cal O}$, are:
\begin{equation}
\delta_S \widehat{\omega}^I = \widehat{\psi}^I, \ \ \ \ \ \ \ \ \delta_S \widehat{\cal S}^I =
\widehat{\Psi}^I,
\end{equation}
\begin{equation}
\delta_S \widehat{\psi}^I_{\mu} = D_{\mu} \widehat{\gamma}^I +  \widehat{\gamma}^{\lambda}*
\partial_{\lambda}
\widehat{\omega}^I_{\mu} + \partial_{\mu}  \widehat{\gamma}^{\lambda} *
\widehat{\omega}^I_{\lambda} \equiv \widehat{\delta}^G_{(\gamma)}
\widehat{\omega}^I_{\mu}
\end{equation}
\begin{equation}
\delta_S \widehat{\Psi}^I=  D_{\mu} \widehat{\gamma}^I + \widehat{\gamma}^{\lambda}*
\partial_{\lambda} \widehat{\cal S}^I + \partial
\widehat{\gamma}^{\lambda} * \widehat{\cal S}^I_{\mu \lambda} \equiv
\widehat{\delta}^G_{(\gamma)}
\widehat{\cal S}^I,
\end{equation}
\begin{equation}
\delta_S \widehat{\gamma}^I=0, \ \ \ \ \ \ \ \ \delta_S \widehat{\gamma}^{\mu}=0,
\label{gamas}
\end{equation}
where $ D_{\mu}\widehat{\gamma}^I = D^{\bf ad}_{\mu} \widehat{\gamma}^I + h^{IJK}
\left[\widehat{\gamma}^J
\stackrel{\ast}{,}\widehat{\omega}^K_{\mu}\right]$ and $\widehat{\delta}^G_{(\gamma)}$ denotes
the Lorentz and diffeomorphism infinitesimal transformation, with transformation parameter given
by $\gamma$.

The anti-ghost fields are: $\widehat{X}_I$ (two-form), $\widehat{\chi}_I$ (one-form) and
$\widehat{\chi}_{IJ}$ (zero-form), while the
auxiliary
fields $\widehat{\Pi}_I$ (two-form),
$\widehat{\pi}_I$ (one-form) and $\widehat{\pi}_{IJ}$ (zero-form) transform as follows:

\begin{equation}
\delta_S \widehat{X}_I = \widehat{\Pi}_I, \ \ \ \ \ \  \delta_S \widehat{\Pi}_I =
\widehat{\delta}^G_{(\gamma)} \widehat{X}_I,
\end{equation}
\begin{equation}
\delta_S \widehat{\chi}_I = \widehat{\pi}_I, \ \ \ \ \ \  \delta_S \widehat{\pi}_I =
 \widehat{\delta}^G_{(\gamma)} \widehat{\chi}_I,
\end{equation}
\begin{equation}
\delta_S \widehat{\chi}_{IJ} = \widehat{\pi}_{IJ}, \ \ \ \ \ \  \delta_S \widehat{\pi}_{IJ} =
 \widehat{\delta}^G_{(\gamma)}\widehat{\chi}_{IJ}.
\end{equation}

Then we propose the $\delta_Q$-invariant Lagrangian

\begin{equation}
L = \delta_Q \int \bigg( \widehat{X}_I \buildrel{*}\over{\wedge}\widehat{\cal R}^I
+ \widehat{\chi}_I \buildrel{*}\over{\wedge}
D\widehat{\cal S}^I
+\frac{1}{2} \widehat{\chi}_{IJ}  \widehat{\cal S}^I \buildrel{*}\over{\wedge} \widehat{\cal
S}^J
\bigg)\nonumber
\end{equation}
\begin{eqnarray}
= &\int& \bigg( \widehat{\Pi}_I \buildrel{*}\over{\wedge} \widehat{\cal R}^I + \widehat{\pi}_I
\buildrel{*}\over{\wedge} D \widehat{\cal S}^I + \frac{1}{2} \widehat{\pi}_{IJ} \widehat{\cal S}^I
\buildrel{*}\over{\wedge} \widehat{\cal S}^J - \widehat{X}_I
\buildrel{*}\over{\wedge} D \widehat{\psi}^I \nonumber\\
&-& \widehat{\chi}_I \buildrel{*}\over{\wedge}
\big(D\widehat{\Psi}^I -
h_{IJK} \widehat{\cal S}^J \buildrel{*}\over{\wedge} \widehat{\psi}^K\big) -
\widehat{\chi}_{IJ} \widehat{\cal S}^I
\buildrel{*}\over{\wedge} \widehat{\Psi}^J \bigg),
\label{hfaction}
\end{eqnarray}
where we have used the BRST transformation proviuosly defined. Here
$\stackrel{\ast}{\wedge}$ is a noncommutative wedge product given, for instance, by:
$\widehat{\bf f} \stackrel{\ast}{\wedge} \widehat{\bf g} = \widehat{f}_{{\mu}_1 \dots
\mu_p}(x) * \widehat{g}_{\nu_1 \dots \nu_q}(x) dx^{{\mu}_1} \wedge \dots \wedge dx^{{\mu}_p}
\wedge dx^{{\nu}_1} \wedge \dots \wedge dx^{{\nu}_q}$. for any $\widehat{\bf f} \in
\Lambda^p(T^*X_{\theta})$ and $\widehat{\bf g} \in \Lambda^q(T^*X_{\theta})$.

The noncommutative gravitational instantons described by the $\widehat{\cal S}_{\cal H}$ structure
are not
independent but, they
are related via the Bianchi identities $D \widehat{\cal R}^I =0$ and $D^2\widehat{\cal S}^I =
h^{IJK} \widehat{\cal R}^J \buildrel{\ast}\over{\wedge}{\cal S}^K.$ This gives rise to the
incorporation of a gauge
symmetry, which mixed with the above preliminary BRST symmetry, leads to the full BRST
symmetry
determined by
the BRST charge $Q$, which, with help of Ref.  \cite{ncbrst}, can be easily shown
that satisfies $Q^2=0$, i.e. $\delta^2_Q \widehat{\cal O} = 0$ for any $\widehat{\cal O}$. This
gauge symmetry is fixed through the incorporation of the ghost multiplet 
$(\widehat{c}^{\mu}, \widehat{\xi}_{IJ},\widehat{\lambda}_I,
\widehat{\eta}_I)$.

For simply connected spacetimes $X_{\theta}$, the noncommutative spin connection 
$\widehat{\omega}^I_{\mu}$ can be gauged away and therefore can be removed from the 
action (\ref{hfaction}). Other fields can also be integrated out from the action
and we have,
\begin{equation}
L'= \int \bigg( \widehat{\pi}_I
\buildrel{*}\over{\wedge} d \widehat{\cal S}^I + \frac{1}{2} \widehat{\pi}_{IJ} \widehat{\cal S}^I
\buildrel{*}\over{\wedge} \widehat{\cal S}^J -  \widehat{\chi}_I \buildrel{*}\over{\wedge}
d\widehat{\Psi}^I  -
\widehat{\chi}_{IJ} \widehat{\cal S}^I
\buildrel{*}\over{\wedge} \widehat{\Psi}^J \bigg),
\end{equation}
with the BRST transformations for the involved fields $(\widehat{\cal S}^I, \widehat{\Psi}^I,
\widehat{\gamma}^{\mu},  \widehat{c}^{\mu}, \widehat{\chi}_{I}, \widehat{\pi}_I,
\widehat{\xi}_{IJ}, \widehat{\pi}_{IJ}, \widehat{\lambda}_I, \widehat{\eta}_I)$  given by: 
\begin{equation}
\delta_Q \widehat{\cal S}^I = \widehat{\Psi}^I + \widehat{\delta}^D_{(c)} \widehat{\cal S}^I, \
\ \ \
\delta_Q \widehat{\Psi}^I = \widehat{\delta}^D_{(\gamma)} \widehat{\cal S}^I +
\widehat{\delta}^D_{(c)}
\widehat{\Psi}^I,
\end{equation}
\begin{equation}
\delta_Q \widehat{\gamma}^{\mu} =
\widehat{c}^{\lambda} * \partial_{\lambda} \widehat{\gamma}^{\mu} -
\widehat{\gamma}^{\lambda} * \partial_{\lambda} \widehat{c}^{\mu}, \ \ \ \
\delta_{Q} \widehat{c}^{\mu} = - \widehat{\gamma}^{\mu} +
\widehat{c}^{\lambda}* \partial_{\lambda} \widehat{c}^{\mu},
\end{equation}
\begin{equation}
\delta_Q \widehat{\chi}_{I} = \widehat{\pi}_{I} + d \widehat{\lambda}_I +
\widehat{\delta}^{D}_{(c)}
\widehat{\chi}_I, \ \ \ \
\delta_Q \widehat{\pi}_I = d \widehat{\eta}_I + \widehat{\delta}^{D}_{(\gamma)} \widehat{\chi}_I
+ \widehat{\delta}^{D}_{(c)} \widehat{\pi}_I,
\end{equation}
\begin{equation}
\delta_Q \widehat{\xi}_{IJ} = \widehat{\pi}_{IJ} + \widehat{\delta}^{D}_{(c)}
\widehat{\chi}_{IJ}, \ \
\ \
\delta_Q \widehat{\pi}_{IJ} =  \widehat{\delta}^{D}_{(\gamma)} \widehat{\chi}_{IJ} +
\widehat{\delta}^{D}_{(c)}
\widehat{\pi}_{IJ},
\end{equation}
\begin{equation}
\delta_Q \widehat{\lambda}_I = -\widehat{\eta}_I +  \widehat{\delta}^{D}_{(c)}
\widehat{\lambda}_I, \ \ \ \
\delta_Q \widehat{\eta}_I = \widehat{\delta}^{D}_{(\gamma)}  \widehat{\lambda}_I +
\widehat{\delta}^{D}_{(c)}
\widehat{\eta}_I.
\end{equation}

The lagrangian $L'$ has associated the partition function, 
\begin{equation}
Z= \int ({\cal D}\Phi) \exp \big( - L'\big),
\end{equation}
where $({\cal D}\Phi)$ abbreviates the measure of all noncommutative fields of the theory.

The BRST invariance of $\delta_Q L'=0$, implies that the partition function $Z$
and the correlation functions of BRST-invariant observables $\widehat{\cal O}$,  {\it i.e.} functionals
of the noncommutative fields contained in $L'$,
\begin{equation}
Z(\widehat{\cal O}) = \langle \widehat{\cal O} \rangle = \int ({\cal D}\Phi) \exp \big( -
L'\big) \cdot \widehat{\cal O},
\end{equation}
are topological invariants and constitute a noncommutative
deformation of the gravitational Donaldson invariants \cite{wtg}. These invariants are
independent on the metric, but they
do depend on the differentiable structure of the noncommutative differentiable manifold
$X_{\theta}$. From inspection of the BRST transformation (\ref{gamas}), the obvious invariants
and their descendents are
those functionals constructed from the
BRST-invariant fields $\widehat{\gamma}^i$ and $\widehat{\gamma}^{\mu}$, such that $\delta_Q
\widehat{\gamma}^i =0$ and $\delta_Q
\widehat{\gamma}^{\mu} =0$.

Moreover, for similar reasons, $Z$ and $Z(\widehat{\cal O})$ are independent of a rescaling
of the volume form, $v\to t v$. Thus, for simplicity, we can evaluate the path
integral for large values of $t$. {\it e.g.} $t \to \infty$. In this limit the path integral is
dominated by the classical minima, and it is now concentrated in the noncommutative instanton
configuration, determined by the following system of equations $\widehat{\cal S}'_{\cal H}$:
\begin{equation}
\delta \widehat{\omega}^I: \ \ \ \ \ \ \  d\widehat{\cal S}^I=0,
\end{equation}
\begin{equation}
\delta \widehat{\cal S}^I: \ \ \ \ \ \ \  \widehat{\cal R}^I=0.
\end{equation}

These equations are the noncommutative $\widehat{\cal S}_{\cal H}$-structure in the ``strong
heaven''
gauge \cite{jerzy}, which determines the moduli space of the heavenly configurations
$\widehat{\cal M}_{\cal H}$. After some
computations, following closely Ref. \cite{jerzy},
we can carry over the above equations into the equivalent form of a
noncommutative deformation of the first and second heavenly equations. After the analysis, the
resulting first heavenly equation is,
\begin{equation}
\partial^2_{pr} \Omega * \partial^2_{qs} \Omega  - \partial^2_{qr} \Omega *  \partial^2_{ps}
\Omega = 1,
\label{fhe}
\end{equation}
which is a nonlineal PDE for the holomorphic function  $\Omega = \Omega(p,q,r,s)$
in $X_{\theta}$, in a local chart $\{x^{\mu}\}=\{p,q,r,s\}$. An equivalent description of the
moduli space $\widehat{\cal M}_{\cal H}$ can be given in the local chart
$\{x^{\mu}\}=\{x,y,p,q\}$ with $x:= \partial_p \Omega$ and  $y:= \partial_q \Omega$. In this
case we get a noncommutative version of the second heavenly equation given by
\begin{equation}
\partial^2_{xx} \Theta * \partial^2_{yy} \Theta - (\partial^2
_{xy}\Theta) * (\partial^2_{xy}
\Theta) + \partial^2_{xp} \Theta +  \partial^2_{yq} \Theta =0,
\label{she}
\end{equation}
for a holomorphic function $\Theta=\Theta(x,y,p,q)$.

%%%%%%%%%%%%%%%%%%%%%%%%%%%%%%%%%%%%%%%%%%%%%%%%%%%%%%%%%%%%%%%%%%%%%%%%%%
%%%%%%%%%%%%%%%%%%%%%%%%%%%%%%%%%%%%%%%%%%%%%%%%%%%%%%%%%%%%%%%%%%%%%%%%%%
\section{Conclusions}
In the present paper, starting from previous results \cite{topologico} concerning the
noncommutative deformation of the classical topological invariants given by the 
Euler number and signature of
a four smooth manifold $X$, we deform the topological half-flat gravity proposed in
Ref. \cite{kuni}. We first
discuss the convenience of having a $SL(2,{\bf C})$ chiral gravity theory, in order to find a
noncommutative Lagrangian for gravity preserving all original symmetries of the theory. An
interesting feature is that this procedure can be implemented to find a noncommutative
deformation of this topological gravity.

Topological half-flat gravity is defined as the intersection theory in the moduli space ${\cal
M}$. This moduli problem becomes important as far as the other topological gravity theories are
cohomological theories of the moduli space of metrics, with an action of the form
(\ref{topoaction}), which involves the Levi-Civita tensor $\varepsilon_{abcd}$, and which seems
to be very difficult to implement in the formulation of a noncommutative theory of gravitation,
without spoiling the full original symmetry.

Topological gravity possesses a fermionic BRST symmetry, which we have noncommutatively deformed
following the procedure of Ref. \cite{ncbrst}. This results in a noncommutative topological
gravity, invariant under the full set of symmetries including the BRST symmetry. For simply
connected spaces $X_{\theta}$, we can gauge away the self-dual spin connection and set
$\widehat{\omega}^I_{\mu}$ to zero. We have shown, by using the stationary phase approximation in
evaluating the partition function, that the main contribution of the Feynman path integral comes
from the moduli space $\widehat{\cal M}_{\cal H}$, characterized as the space of solutions of the
noncommutative formulation of the heavenly equations (\ref{fhe}) or (\ref{she}). Thus in
noncommutative topological half-flat gravity, the gravitational analogs of Donaldson invariants
can be computed as the intersection form in $\widehat{\cal M}_{\cal H}$ determined by the
heavenly equation. A detailed analysis of this result is left for future work.

%\vskip 2truecm
%%%%%%%%%%%%%%%%%%%%%%%%%%%%%%%%%%%%%%%%%%%%%%%%%%%%%%%%%%%%%%%%%%%%%%%%%%
%%%%%%%%%%%%%%%%%%%%%%%%%%%%%%%%%%%%%%%%%%%%%%%%%%%%%%%%%%%%%%%%%%%%%%%%%%
\centerline{\bf Acknowledgments}
This work was supported in part by CONACyT M\'exico Grants Nos. 37851E and
33951E.

%%%%%%%%%%%%%%%%%%%%%%%%%%%%%%%%%%%%%%%%%%%%%%%%%%%%%%%%%%%%%%%%%%%%%%%%%%%%

%\vskip 2truecm
%%%%%%%%%%%%%%%%%%%%%%%%%%%%%%%%%%%%%%%%%%%%%%%%%%%%%%%%%%%%%%%%%%%%%%%
%%%%%%%%%%%%%%%%%%%%%%%%%%%%%%%%%%%%%%%%%%%%%%%%%%%%%%%%%%%%%%%%%%%%%%%

%\begin{references}


\begin{thebibliography}{99}

\bibitem{connesbook}  A. Connes, {\it Noncommutative Geometry}, Academic Press
(1994).

\bibitem{cham1}A.H. Chamseddine, {Commun. Math. Phys.} {\bf 218}(2001) 283.

\bibitem{moffat}  J.W. Moffat, {Phys. Lett.} B {\bf 491} (2000) 345; {Phys.
Lett.} B {\bf 493} (2000) 142.

\bibitem{cham}  A.H. Chamseddine, {Phys. Lett.} B {\bf 504} (2001) 33.

\bibitem{cham2}  A.H. Chamseddine, ``Invariant Actions for Noncommutative
Gravity'', hep-th/0202137.

\bibitem{zanon} M.A. Cardella and Daniela Zanon, ``Noncommutative Deformation of
Four-dimensional Einstein Gravity", hep-th/0212071.

\bibitem{wess5} B. Jurco, L. Moller, S. Schraml, P. Schupp and J. Wess, {Eur.
Phys. J. C} {\bf 21} (2001) 383.

\bibitem{topologico} H. Garc\'{\i}a-Compe\'an, O. Obreg\'{o}n, C. Ram\'{\i}rez, and M. Sabido,
Phys. Rev. D. {\bf 68} (2003) 045010, hep-th/0210203.

\bibitem{reviewsdg} P. Peld\'an, ``Actions for Gravity, with Generalizations: a Review", Class.
Quant. Grav. {\bf 11} (1994) 1087, gr-qc/9305011; S. Carlip, ``Quantum Gravity: a Progress Report",
Rept. Prog. Phys. {\bf 64}
(2001) 885, gr-qc/0108040; G.T. Horowitz, ``Quantum Gravity at the Turn of the Millennium",
gr-qc/0011089.

\bibitem{ashtekar} A. Ashtekar, Phys. Rev. Lett. {\bf 77} (1986) 3228; Phys. Rev. D {\bf 36}
(1987) 1587; {\it Lectures on Non-perturbative Canonical Gravity}, World Scientific, Singapore
(1991).

\bibitem{ncsdg} H. Garc\'{\i}a-Compe\'an, O. Obreg\'{o}n, C. Ram\'{\i}rez, and M. Sabido, Phys.
Rev. D. {\bf 68} (2003) 044015, hep-th/0302180.

\bibitem{pleban} J. Pleba\'nski, J. Math. Phys. {\bf 18} (1977) 2511.

\bibitem{helicity} J. Pleba\'nski, ``The Spinorial and Helicity Formalisms of Riemannian
Structures in Complex or Real Four Dimensions'', Lecture Notes, CINVESTAV 1985.

\bibitem{gary} G.T. Horowitz, Commun. Math. Phys. {\bf 125} (1989) 417; G.T. Horowitz and M.
Srednicki, Commun. Math. Phys. {\bf 130} (1990) 83.

\bibitem{kuni} H. Kunitomo, ``Topological Half-flat Gravity'', Mod. Phys. Lett. A {\bf 6}
(1991) 2389.

\bibitem{jerzy} J. Pleba\'nski, J. Math. Phys. {\bf 16} (1975) 2395.

\bibitem{topoym} E. Witten, ``Topological Quantum Field Theories'', Commun. Math. Phys.
{\bf 117} (1988) 117.

\bibitem{wtg} E. Witten, ``Topological Gravity'', Phys. Lett. B {\bf 206} (1988) 601.

\bibitem{brooks} R. Brooks, D. Montano and J. Sonnenschein,  Phys. Lett. B {\bf 214} (1988) 91.

\bibitem{labas} J.M.F. Labastida and M. Pernici, ``A Lagrangian for Topological Gravity and
its
BRST Quantization'', Phys. Lett. B {\bf 213} (1988) 319.

\bibitem{my} R. Myers and V. Periwal, ``Invariants of Smooth 4-Manifolds From Topological
Gravity'', Nucl. Phys. B {\bf 361} (1991) 290.

\bibitem{teo} M.J. Perry and E. Teo, ``Topological Conformal Gravity in Four Dimensions'', Nucl.
Phys. B {\bf 401} (1993) 206.

\bibitem{tgi} A. Nakamichi, I. Oda and A. Sugamoto, Phys. Rev. D {\bf 44} (1991) 3835; I. Oda and
A. Sugamoto, ``Topological Four-dimensional Einstein Gravity'',
Phys. Lett. B {\bf 266} (1991) 280.

\bibitem{ncbrst} M. Soroush, ``BRST Quantization of Noncommutative Gauge Theories'', Phys. Rev.
D {\bf 67} (2003) 105005-1.



%\end{references}
\end{thebibliography}
\end{document}